\def\upd{{\rm d}}

\documentclass[pre,showpacs]{revtex4}
\usepackage{graphicx}

\begin{document}

\title{Exact solution of a Brownian inchworm model for self-propulsion}

\author{A. Baule}
\email{abaule@rockefeller.edu.}

\affiliation{The Rockefeller University, 1230 York Avenue, New York, NY 10065, USA}

\author{K. Vijay Kumar}
\email{vijayk@physics.iisc.ernet.in}
\affiliation{ CCMT, Department of Physics,
Indian
Institute of Science, Bangalore 560012, India}

\author{Sriram Ramaswamy}
\email{sriram@physics.iisc.ernet.in}
\altaffiliation[Also at~]{CMTU, JNCASR, Bangalore 560064, India.}
\affiliation{ CCMT, Department of Physics,
Indian Institute of Science, Bangalore 560012, India}

\begin{abstract}

We present the exact solution of a Brownian inchworm model of a self-propelled
elastic dimer which has recently been proposed in  [K. V. Kumar \textit{et al},
Phys. Rev. E \textbf{77}, 020102(R) (2008)] as a unifying model for the
propulsion mechanisms of DNA helicase, polar rods on a vibrated surface,
crawling keratocytes, and Myosin VI.

\end{abstract}


\maketitle

\section{Introduction}

Methods of nonequilibrium statistical mechanics are particularly useful in describing transport processes in biological systems. In this article we investigate a nonequilibrium model of a molecular motor (or self-propelled particle), in which unidirectional movement is generated by rectification of Brownian diffusive motion. A variety of dynamical processes in cells arise from motor proteins such as kinesin, myosin or DNA helicase, which are driven by hydrolysis of ATP into ADP and move along tracks consisting of microtubules, actin filaments, or DNA respectively \cite{Howard,Yu06,Altman04}. In larger assemblies molecular motors give rise to motion of bacteria and are ultimately responsible for muscle contraction and macroscopic movement of living organisms. But such directed motion is also observable in the non-living world, for example in agitated granular matter \cite{Yamada03,Dorbolo05}. From a physicist's point of view, one of the main interests in these self-propelled systems concerns the basic principles by which directed motion is generated from isotropic energy input in the absence of an externally imposed gradient.

In the above mentioned systems the unidirectional movement of the center of
mass results quite generically from a coupling to the internal coordinates
under the influence of (i) energy input due to chemical or mechanical
nonequilibrium noise and (ii) an asymmetric environment for the internal
coordinates. A widely used model of a Brownian motor implementing these
principles is based on an asymmetric ratchet potential, acting on the centre 
of mass of the particle \cite{julicheretal}. Models of this
type date back to pioneering work of Smoluchowski, Feynman, and Huxley (see
\cite{Reimann02,Kolomeisky07} and references therein), and have subsequently attracted a lot
of research activities. The basic concept is that of a Brownian particle moving
in an asymmetric periodic potential. Due to the second law of thermodynamics,
the asymmetric energy landscape and thermal noise alone are not sufficient to
generate directed motion. A common way to model the effect of the required
nonequilibrium energy input is to assume that it gives rise to a rapid increase
in the system temperature thus effectively resulting in a periodic switching
between high and low temperature states. At high temperature the particle can
freely diffuse over the potential barriers, whereas at low temperature it will
be trapped in a potential well. Due to the spatial asymmetry a periodic
switching between these states will cause a net movement in a direction
prescribed by the asymmetry. Such ratchet models might serve as crude
simplifications of the movement mechanism of motor proteins, where the
asymmetric periodic potential results from the interaction of the motor with
the track and ATP hydrolysis is the cause of the sudden temperature jumps.

An alternative picture of a Brownian motor can be established by modelling the motor-track interaction in terms of an effective friction force \cite{Mogilner98}. Very recently an inchworm model of a self-propelled particle has been proposed in \cite{Kumar08}, where the required spatial asymmetry is implemented by such a friction force. This model consists of two beads connected by an elastic spring, which are driven by thermal and nonequilibrium noise. Each bead experiences a different, stretch dependent friction, which is sufficient to generate a non-zero center of mass velocity in a fixed direction. On the basis of this new rectification mechanism the propulsion in systems as diverse as polar rods on a vibrating surface, DNA helicase, and Myosin VI on actin, can be understood in a unified way, as discussed in \cite{Kumar08}.

The purpose of the present paper is to further investigate the new Brownian inchworm model of Ref.~\cite{Kumar08}, which so far has only been treated perturbatively and numerically. In particular, we present exact solutions for the average center of mass velocity and the distribution of internal coordinate of the inchworm in the overdamped regime. The exact solution allows for a strict derivation of some of the perturbative results and shows excellent agreement with simulation data. Furthermore, the effect of a more realistic spring potential, such as the FENE spring, is investigated.

\section{The Brownian inchworm}
\label{I_Sec_BI}

The Brownian inchworm is described in terms of two coupled Langevin equations for the positions of the two beads $x_1$ and $x_2$ under the influence of thermal and nonequilibrium noise \cite{Kumar08}:
\begin{eqnarray}
\label{motion_eq}
m_i\ddot{x}_i(t)+\gamma_i(x)\dot{x}_i(t)=-\partial_i U(x)+\sqrt{2\gamma_i(x)k_BT}\eta_i(t)+\sqrt{A_i}\zeta_i(t).
\end{eqnarray}
Here, $\partial_i\equiv\partial/\partial x_i$ with $i\in\{1,2\}$ and $U(x)$ is the spring potential depending on the relative coordinate $x\equiv x_1-x_2-x_0$ for an equilibrium extension $x_0$. The crucial ingredient in this model is the stretch dependent friction $\gamma_i(x)\geq0$, acting independently on the two beads. The thermal noise $\eta_i(t)$ has Gaussian characteristics with zero mean $\left<\eta_i(t)\right>=0$ and correlation $\left<\eta_i(t)\eta_j(t')\right>=\delta_{ij}\delta(t-t')$. The thermal character is expressed in the prefactor $\sqrt{2\gamma_i(x)k_BT}$ of $\eta_i(t)$, stemming from the fluctuation dissipation relation. Likewise, we assume Gaussian statistics with zero mean and delta-correlation for the nonequilibrium noise $\zeta_i(t)$, which, in contrast to the thermal noise, does not satisfy a fluctuation-dissipation relation as it is considered to represent the external nonequilibrium energy input. 

The two uncorrelated Gaussian processes in Eq.~(\ref{motion_eq}) can be superposed into a single one, denoted by $\xi_i(t)$, whereby $\left<\xi_i(t)\right>=0$ and $\left<\xi_i(t)\xi_j(t')\right>=\delta_{ij}\delta(t-t')$. The equations of motion then read
\begin{eqnarray}
\label{motion_eq2}
m_i\ddot{x}_i(t)+\gamma_i(x)\dot{x}_i(t)=-\partial_iU(x)+g_i(x)\xi_i(t),
\end{eqnarray}   
where
\begin{eqnarray}
\label{subs_g}
g_i(x)\equiv\sqrt{2\gamma_i(x)k_BT+A_i}.
\end{eqnarray}

Important time scales in this system are the relaxation time of the center of mass velocity and of the fast and slow modes of the relative coordinate. When the center of mass relaxes much faster than the relative coordinate, the system is effectively in the overdamped regime and the influence of inertia can be neglected.
Independent of the precise values of the timescales, it is clear that, 
in a system where dissipation dominates, the correlation time of the noise 
is the shortest, that of the velocities is next ($\sim m/\gamma$), and 
that of the relative coordinate is the longest ($\sim \gamma/\kappa$) where 
$\kappa$ is the spring constant, $\gamma$ is a typical value of the $x$-dependent 
damping, and $m$ is a typical mass. Thus one is always in the regime 
where the adiabatic elimination procedure can be done with a white noise 
approximation. 

\begin{figure}
\begin{center}
\includegraphics[width=6cm]{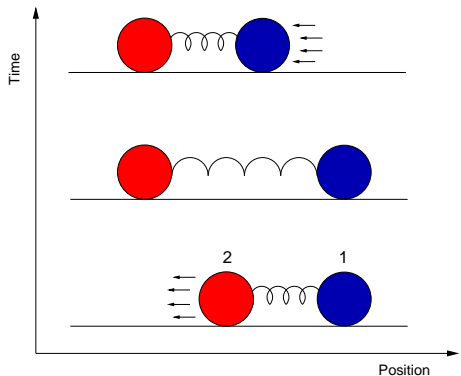} 
\caption{\label{I_Fig_inch}Depiction of the Brownian inchworm. A movement as shown here can be observed, for example, if (i) the friction on bead $2$ is constant and stretch-dependent on bead $1$ (increases for larger stretch), and (ii) the nonequilibrium noise-strengths are equal.}
\end{center}
\end{figure}

In the steady-state regime the Brownian inchworm exhibits unidirectional movement which can be described as follows. Let us consider the case of constant friction on bead $2$ while bead $1$ experiences stretch-dependent damping (increasing for larger stretch). Gaussian noise results in a high frequency of stochastic kicks compared with the relaxation rate of the relative coordinate, such that the dimer basically never relaxes while being under the influence of the noise. Since bead $2$ has a higher mobility it will accumulate a larger displacement than bead $1$ whose movement is inhibited with increased stretch. Ultimately bead $1$ follows bead $2$ since this direction is favoured due to the friction and the spring force. In this case the net effect of the fluctuations is a movement of the center of mass in the negative direction (see Fig.~\ref{I_Fig_inch}).

However, both center of mass velocity and direction of movement depend on the magnitudes of the noise strengths $A_1$ and $A_2$. For $A_2>A_1$ a current reversal can be observed. We find quite generally that, whenever $A_2\neq A_1$, the direction of movement is dominated by the asymmetric noise input rather than by the asymmetric friction. This is further discussed in Sec.~\ref{I_Sec_asym} below.

\section{Exact solution in the overdamped regime}

An analytical treatment is accessible in the overdamped regime where the dynamics exhibits a time scale separation between position (slow) and velocity (fast) degrees of freedom. It is important to note that the position-dependent friction in the
equations of motion (\ref{motion_eq2}) requires a careful inspection of the adiabatic
elimination \cite{VanKampen2}. See \cite{Gardiner,Ryter} for a
discussion of the issues involved. At least for white noise sources,
the mathematically sound way \cite{Ryter} to carry out the elimination
of the velocity is to start with the Fokker-Planck equation for the
distribution of position and velocity and extract the Smoluchowski
equation for the distribution for position alone, through an expansion
in $1/\gamma$. Working backwards from the resulting Smoluchowski
equation yields the overdamped Langevin equation \cite{Ryter,Sancho82}
for the position coordinates $x_i$ (see Appendix~\ref{appendix})
\begin{eqnarray}
\label{OD_correct}
\dot{x}_i &=& -\frac{\partial_iU}{\gamma_i(x)}-\frac{1}{2}\frac{g_i(x)\partial_ig_i(x)}{\gamma_i(x)^2}+\frac{g_i(x)}{\gamma_i(x)}\xi_i(t),
\end{eqnarray}
with multiplicative noise interpreted \textit{\`{a} la} Stratonovich.

We will see below that Eq.~(\ref{OD_correct}) with Stratonovich interpretation gives rise to the correct equilibrium distribution for the relative coordinate $x$, namely $p_{st}(x)\propto e^{-\beta U(x)}$, in the absence of the active noise. If one wants to attribute a different interpretation to the multiplicative noise terms, an additional drift term has to be added in Eq.~(\ref{OD_correct}) in order to obtain the physically correct distribution for $p_{st}(x)$. There exists in fact a whole one parameter family of overdamped Langevin equations which are equivalent to Eq.~(\ref{OD_correct}) \cite{Lau07}.

In the following the prime denotes a derivative with respect to the relative coordinate $x$, so that $\partial_1 U(x)=-\partial_2 U(x)=U(x)'$. With Eq.~(\ref{OD_correct}) the equation of motion for the center of mass coordinate can be written as
\begin{eqnarray}
\label{motion_eq_cm}
\dot{x}_{cm}(t)&=&-\frac{1}{2}\left(\frac{1}{\gamma_1(x)}-\frac{1}{\gamma_2(x)}\right)U'(x)-\frac{1}{4}\left(\frac{g_1(x)g_1'(x)}{\gamma_1(x)^2}-\frac{g_2(x)g_2'(x)}{\gamma_2(x)^2}\right)\nonumber\\
&&+\frac{g_1(x)}{2\gamma_1(x)}\xi_1(t)+\frac{g_2(x)}{2\gamma_2(x)}\xi_2(t).
\end{eqnarray}
Likewise, the equation of motion for the relative coordinate reads
\begin{eqnarray}
\label{motion_eq_r}
\dot{x}(t)&=&-\left(\frac{1}{\gamma_1(x)}+\frac{1}{\gamma_2(x)}\right) U'(x)-\frac{1}{2}\left(\frac{g_1(x)g_1'(x)}{\gamma_1(x)^2}+\frac{g_2(x)g_2'(x)}{\gamma_2(x)^2}\right)\nonumber\\
&&+\frac{g_1(x)}{\gamma_1(x)}\xi_1(t)-\frac{g_2(x)}{\gamma_2(x)}\xi_2(t).
\end{eqnarray}
The key mathematical observations here are: (i) the equation of motion for the center of mass is a function of the relative coordinate and the noise only. (ii) The equation of motion for the relative coordinate is independent of the center of mass coordinate. As a consequence of these properties we obtain the average center of mass velocity directly by averaging Eq.~(\ref{motion_eq_cm}) over the stochastic realizations
\begin{eqnarray}
\label{I_sol1}
\left<\dot{x}_{cm}\right>&=&-\frac{1}{2}\left<\left(\frac{1}{\gamma_1(x)}-\frac{1}{\gamma_2(x)}\right)U'(x)\right>-\frac{1}{4}\left<\left(\frac{g_1(x)g_1'(x)}{\gamma_1(x)^2}-\frac{g_2(x)g_2'(x)}{\gamma_2(x)^2}\right)\right>\nonumber\\
&&+\frac{1}{2}\left<\frac{g_1(x)}{\gamma_1(x)}\xi_1(t)\right>+\frac{1}{2}\left<\frac{g_2(x)}{\gamma_2(x)}\xi_2(t)\right>.
\end{eqnarray}
It is important to note that the contributions of the multiplicative noise terms are non-zero due to the Stratonovich interpretation and can be calculated with the help of Novikov's theorem for Gaussian processes.

For an arbitrary functional $u[\eta(t)]$ of a delta-correlated zero mean Gaussian process $\eta(t)$, this theorem states that the average over the product $\left< u[\eta(t)]\eta(t)\right>$ can be expressed as an average over the response of $u$ to a change in $\eta(t)$ \cite{Novikov65,Fox86}
\begin{eqnarray}
\left< u[\eta(t)]\eta(t)\right>=\left<\frac{\delta u[\eta(t)]}{\delta \eta(t)}\right>.
\end{eqnarray}
In the present case, the product rule for the functional derivative leads to
\begin{eqnarray}
\left<\frac{g_i(x)}{\gamma_i(x)}\xi_i(t)\right>=\left<\left(\frac{g_i(x)}{\gamma_i(x)}\right)'\frac{\delta x(t)}{\delta \xi_i(t)}\right>.
\end{eqnarray}
The response function $\delta x(t)/\delta \xi_i(t)$ is calculated from Eq.~(\ref{motion_eq_r}) as \cite{Fox86}
\begin{eqnarray}
\label{I_fderiv}
\frac{\delta x(t)}{\delta \xi_i(t)}=\pm\frac{1}{2} \frac{g_i(x)}{\gamma_i(x)},
\end{eqnarray}
where the $+$ sign applies to $i=1$ and the $-$ sign to $i=2$ since the relative coordinate has been defined as $x=x_1-x_2-x_0$. The factor $1/2$ is due to the Stratonovich interpretation. The final result for the average over the multiplicative noise terms is 
\begin{eqnarray}
\frac{1}{2}\left<\frac{g_i(x)}{\gamma_i(x)}\xi_i\right>=\pm\frac{1}{4}\left<\left(\frac{g_i(x)g'_i(x)}{\gamma_i(x)^2}-\frac{g_i(x)^2\gamma_i(x)'}{\gamma_i(x)^3}\right)\right>,
\end{eqnarray}
which can be substituted into Eq.~(\ref{I_sol1}). The average center of mass velocity is then
\begin{eqnarray}
\left<\dot{x}_{cm}\right>&=&-\frac{1}{2}\left<\left(\frac{1}{\gamma_1(x)}-\frac{1}{\gamma_2(x)}\right)U(x)'\right>-\frac{1}{4}\left<\frac{g_1(x)^2\gamma_1(x)'}{\gamma_1(x)^3}\right>+\frac{1}{4}\left<\frac{g_2(x)^2\gamma_2(x)'}{\gamma_2(x)^3}\right>.\nonumber\\
\end{eqnarray}
Substitution of the $g_i$, Eq.~(\ref{subs_g}), yields our first main result, an exact expression for the average inchworm velocity in the overdamped regime:
\begin{eqnarray}
\label{solution}
\left<\dot{x}_{cm}\right>&=&-\frac{1}{2}\left<\left(\frac{1}{\gamma_1(x)}-\frac{1}{\gamma_2(x)}\right)U(x)'\right>+\frac{k_BT}{2}\left<\left(\frac{1}{\gamma_1(x)}-\frac{1}{\gamma_2(x)}\right)'\right>\nonumber\\
&&+\frac{A_2}{4}\left<\frac{\gamma_2(x)'}{\gamma_2(x)^3}\right>-\frac{A_1}{4}\left<\frac{\gamma_1(x)'}{\gamma_1(x)^3}\right>.
\end{eqnarray}

The averages can be performed if the distribution of the relative coordinate $x$ is known. Since the equation of motion for $x$, Eq.~(\ref{motion_eq_r}), is decoupled from the center of mass coordinate, this distribution is determined by solving the Fokker-Planck equation corresponding to the Langevin Eq.~(\ref{motion_eq_r}). In order to simplify notation let us rewrite Eq.~(\ref{motion_eq_r}) as
\begin{eqnarray}
\label{motion_eq_r_simple}
\dot{x}=a(x)+b(x)\xi(t),
\end{eqnarray}
where $\xi(t)$ is the superposition of $\xi_1(t)$ and $\xi_2(t)$ with the same Gaussian white statistics and, upon substitution of the $g_i$, the auxiliary functions $a(x)$ and $b(x)$ are defined as
\begin{eqnarray}
a(x)&\equiv& -\left(\frac{1}{\gamma_1(x)}+\frac{1}{\gamma_2(x)}\right)U'(x)+\frac{k_BT}{2}\left(\frac{1}{\gamma_1(x)}+\frac{1}{\gamma_2(x)}\right)'\\
b(x)&\equiv&\sqrt{2k_BT\left(\frac{1}{\gamma_1(x)}+\frac{1}{\gamma_2(x)}\right)+\frac{A_1}{\gamma_1(x)^2}+\frac{A_2}{\gamma_2(x)^2}}.
\end{eqnarray}
The Fokker-Planck equation for the probability distribution $p(x,t)$ associated with Eq. (\ref{motion_eq_r_simple}) reads in Stratonovich interpretation \cite{Risken}
\begin{eqnarray}
\label{I_xdist_FP}
\frac{\partial}{\partial t}p(x,t)=-\frac{\partial}{\partial x}a(x)p(x,t)+\frac{1}{2}\frac{\partial}{\partial x}b(x)\frac{\partial}{\partial x}b(x)p(x,t).
\end{eqnarray}
If the averages in Eq.~(\ref{solution}) are calculated with $p(x,t)$ as solution of Eq.~(\ref{I_xdist_FP}), one obtains the full time-dependence of the average inchworm velocity. However, apart from the fact that the time-dependent solution is usually very difficult to find, we are also mainly interested in the steady state properties of the inchworm, where the center of mass is expected to move with a constant velocity. The solution in the stationary regime is determined in a straightforward way by setting $\partial_tp(x,t)=0$. Additionally, the probability flux can be set to zero since the relative coordinate is bounded in the spring potential. This simplifies the Fokker-Planck equation to
\begin{eqnarray}
-\frac{2a(x)}{b(x)}p(x)+\frac{\partial}{\partial x}b(x)p(x)=0.
\end{eqnarray}
The solution is then found by straightforward integration and reads
\begin{eqnarray}
\label{distribution}
p(x)=N\frac{1}{b(x)}\exp\left\{\int^x\frac{2a(y)}{b(y)^2}\upd y\right\},
\end{eqnarray}
where $N$ is the normalization constant. Eq.~(\ref{distribution}) is our second main result, an exact expression for the stationary probability distribution of the relative coordinate. With this distribution we can explicitly calculate the average velocity of the inchworm in the steady state given by Eq.~(\ref{solution}).

\section{Comparison with simulation data}
\label{I_Sec_sim}

In this section we compare the exact solution for the average inchworm velocity, Eq.~(\ref{solution}), with the results from a direct simulation of the inchworm. In the simulation the coupled Langevin equations~(\ref{motion_eq2}) retaining inertia have been numerically time-stepped in the overdamped regime using an Euler-Maruyama scheme \textbf{\cite{kloeden}}. An asymmetric damping is assumed, where for simplicity the friction acting on bead $2$ is set constant and the friction on bead $1$ is chosen in the form of (see \cite{Kumar08})
\begin{eqnarray}
\label{I_fric_vj}
\gamma_1(x)=1+\gamma_0\,w\,\tanh(x/w),
\end{eqnarray}
which prescribes a friction varying between two extremal values. The difference between these values is given by $2w\gamma_0$, where $w$ parametrizes the width of the crossover regime and the sign of $\gamma_0$ specifies the orientation. For positive $\gamma_0$ the damping increases for larger $x$ and for negative $\gamma_0$ it decreases.

\begin{figure}
\begin{center}
\includegraphics[width=12cm]{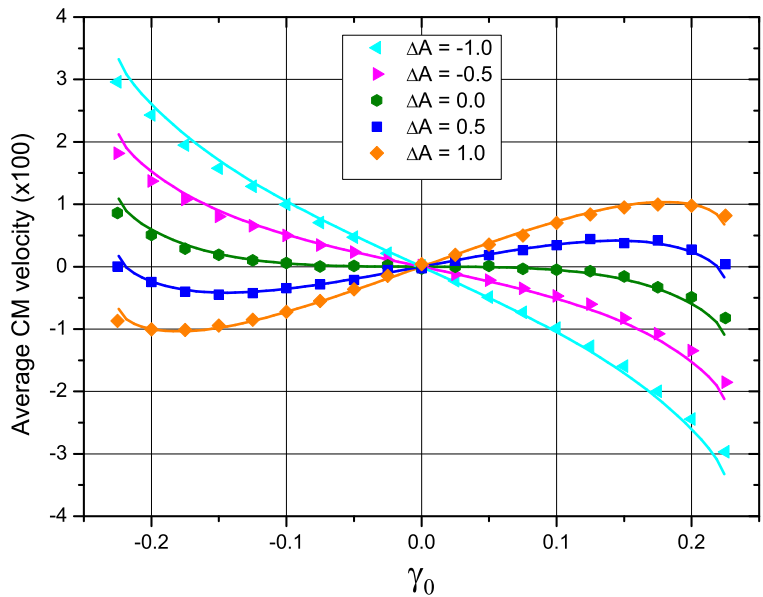}
\caption[The theoretical prediction Eq.~(\ref{solution}) compared with simulation data]{\label{Fig_CMvec}The theoretical prediction Eq.~(\ref{solution}) compared with the simulation data of Fig.~1 in \cite{Kumar08}. We define $\Delta A\equiv A_2-A_1$, where the noise strength $A_1$ is fixed at $1.0$ and $A_2$ is varied for the different curves. Note that in our notation $\gamma_0$ is $\gamma_1$ of \cite{Kumar08}. Parameter values: $\kappa=0.05$, $k_BT=0.01$, $w=4.0$.}
\end{center}
\end{figure}

For the parabolic spring potential $U(x)=\frac{1}{2}\kappa x^2$ and friction term Eq.~(\ref{I_fric_vj}) the distribution $p(x)$, Eq.~(\ref{distribution}), has a lengthy but exact analytical form. With this distribution the averages in Eq.~(\ref{solution}) have been calculated by numerical integration. Fig.~\ref{Fig_CMvec} shows a comparison of the exact solution with the simulation results of Fig.~1 in \cite{Kumar08}. Theory and simulation show excellent agreement such that error bars have been omitted. For $|\gamma_0|>0.22$ a systematic deviation can be observed which is due to the fact that for the given form of the friction term, Eq.~(\ref{I_fric_vj}) with $w=4.0$, $1/\gamma_1(x)$ becomes singular if $|\gamma_0|$ approaches $0.25$. This leads to an increased error in the numerical integration.

As discussed in \cite{Kumar08} an interesting feature in Fig.~\ref{Fig_CMvec} is the occurrence of current reversals for a variation of the noise strengths on each bead. In the case $A_1=A_2$ the inchworm moves due to the asymmetric friction in the direction of the bead with constant damping, in agreement with the propulsion mechanism explained in Sec.~\ref{I_Sec_BI}. The direction of movement is reversed if the noise strength on bead $2$ is increased. In that case the asymmetry of the noise input rather than the asymmetry of the friction dominates the movement. This becomes clearer below when the case of equal stretch dependent friction on both beads is considered.

\begin{figure}
\begin{center}
\includegraphics[width=8cm]{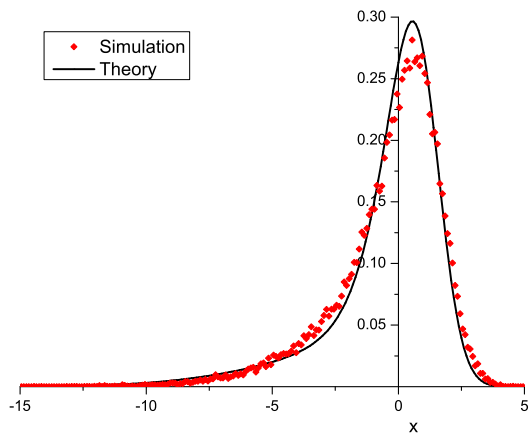}
\caption{\label{Fig_Distribution}The exact expression for $p(x)$ Eq.~(\ref{distribution}) compared with the distribution sampled from simulation data for small but finite inertia. Parameter values: $\kappa=0.1$, $T=0$, $A_2=0$, $\gamma_1=0.4$, $w=2.0$.}
\end{center}
\end{figure}

For another set of parameter values the distribution of the relative coordinate has been determined from the simulation and agrees well with the exact overdamped result Eq.~(\ref{distribution}) (see Fig.~\ref{Fig_Distribution}).   

\section{Special cases of the exact solution}
\label{I_Sec_comp}

Having obtained an exact solution for the average inchworm velocity, it is possible to verify some of the perturbative results in \cite{Kumar08}. We focus on relevant limit cases of the solution and highlight the importance of the nonequilibrium noise as well as the asymmetry and the stretch dependent friction for the movement of the dimer.
\begin{enumerate}

\item On physical grounds it is obvious that for equal and opposite forces on the two beads no net movement should result. From the exact solution it is clear that for $\gamma_1(x)=\gamma_2(x)$ and $A_1=A_2$ all terms in Eq.~(\ref{solution}) vanish pairwise and therefore $\left<\dot{x}_{cm}\right>=0$.
	
\item In the absence of nonequilibrium driving the distribution is required to assume the equilibrium Boltzmann form and due to the second law of thermodynamics the average inchworm velocity should be zero. This can be seen as follows. For $A_1=A_2=0$ the distribution of $x$, Eq.~(\ref{distribution}), reads
\begin{eqnarray}
p(x)&=&N\frac{1}{\sqrt{2k_BT\left(\gamma_1(x)^{-1}+\gamma_2(x)^{-1}\right)}}\exp\left\{\int^x\left(-\frac{U'(y)}{k_BT}+\frac{1}{2}\frac{\left(\gamma_1(y)^{-1}+\gamma_2(y)^{-1}\right)'}{\gamma_1(y)^{-1}+\gamma_2(y)^{-1}}\right)\upd y\right\}.
\end{eqnarray}
Both integrals in the exponent are easily performed. Noting that the second integral yields a logarithmic term which subsequently cancels with the prefactor, the Boltzmann distribution $p(x)\propto e^{-\beta U(x)}$ is readily recovered. Using the equilibrium distribution the first term in Eq.~(\ref{solution}) can be rewritten as 
\begin{eqnarray}
-\frac{1}{2}\left<\left(\frac{1}{\gamma_1(x)}-\frac{1}{\gamma_2(x)}\right)U(x)'\right>&=&\frac{k_BT}{2}\int_{-\infty}^\infty\left(\frac{1}{\gamma_1(x)}-\frac{1}{\gamma_2(x)}\right)\frac{\upd}{\upd x}p(x)\upd x\nonumber\\
&=&-\frac{k_BT}{2}\left<\left(\frac{1}{\gamma_1(x)}-\frac{1}{\gamma_2(x)}\right)'\right>,
\end{eqnarray}
where partial integration has been used in the last step. Therefore the remaining two terms in the expression for the average velocity cancel and we see that without nonequilibrium driving as expected $\left<\dot{x}_{cm}\right>=0$.

\item The crucial ingredient in the inchworm model for the rectification of the diffusive motion is the stretch dependent damping. This is evident from the exact solution. For $\gamma_1$ and $\gamma_2$ both independent of $x$ the distribution of $x$ is $p(x)\propto\exp\left\{-CU(x)/k_BT\right\}$, with a constant $C$. The average center of mass velocity is then given as
\begin{eqnarray}
\left<\dot{x}_{cm}\right>&\propto& \left< U'(x)\right>\propto\int_{-\infty}^{\infty}\frac{\upd}{\upd x}p(x)\upd x=0.
\end{eqnarray}
This confirms that stretch dependent damping is required for directed motion of the inchworm. However, the asymmetry in the system does not necessarily have to originate from an asymmetric damping as discussed in the following.

\end{enumerate}

\subsection{Asymmetric noise strengths}
\label{I_Sec_asym}

In order to further investigate the influence of asymmetric nonequilibrium noise, we consider the special case of equal stretch-dependent friction on both beads (`symmetric friction'), $\gamma_1(x)=\gamma_2(x)$, but asymmetric noise strengths $A_1\neq A_2$. In this case the distribution of the relative coordinate can be written as
\begin{eqnarray}
p(x)=\frac{N\gamma_1(x)}{\sqrt{4k_BT\gamma_1(x)+A_1+A_2}}\exp\left\{-\int^x\frac{4U'(y)\gamma_1(y)+2k_BT\gamma_1'(y)}{4k_BT\gamma_1(y)+A_1+A_2}\upd y\right\}.
\end{eqnarray}
The second integral yields a logarithmic term which can be absorbed into the prefactor. The result is
\begin{eqnarray}
\label{I_dist_asym}
p(x)\propto\frac{\gamma_1(x)}{\gamma_1(x)+(A_1+A_2)/(4k_BT)}\exp\left\{-\int^x\frac{U'(y)\gamma_1(y)}{k_BT\gamma_1(y)+(A_1+A_2)/4}\upd y\right\}.
\end{eqnarray}
In turn, the average inchworm velocity is obtained directly from Eq.~(\ref{solution}) as:
\begin{eqnarray}
\label{I_av_asym}
\left<\dot{x}_{cm}\right>=\frac{1}{4}\left<\frac{\gamma_1'(x)}{\gamma_1(x)^3}\right>(A_2-A_1),
\end{eqnarray}
which is the exact analogue of the perturbative result Eq.~(3) in \cite{Kumar08}. This becomes evident when we expand $\gamma_1(x)$ in powers of $x$ and truncate after the zeroth order.

Assuming that the average in Eq.~(\ref{I_av_asym}) performed with the distribution Eq.~(\ref{I_dist_asym}) is generally non-zero, we therefore establish that asymmetric driving in combination with stretch-dependent damping generates unidirectional center of mass movement. If we consider a fixed total noise strength $A_1+A_2$ and vary the strength on each bead, we see that the distribution $p(x)$ remains unchanged while the average inchworm velocity is exactly linearly proportional to the difference in noise strengths. It is then immediately evident that a sign change of $A_2-A_1$ leads to a reversal of the direction of motion. This is a universal result for symmetric stretch dependent friction and independent of the particular form of the spring potential.

A difference in the active noise on the two heads should be realisable
experimentally in artifical systems. If we generalize the ``Janus beads'' of
\cite{ayushmansen,ramin,kapral} by connecting two catalyst-coated beads of different size by a flexible polymer, we should obtain a self-propelling dimer with asymmetric active noise.

\subsection{FENE spring potential}

In Eq.~(\ref{I_av_asym}) it is indicated that with a continuous increase of one of the $A_i$ the velocity of the inchworm is also increased indefinitely. By contrast, a real molecular motor cannot hydrolyse an arbitrary amount of ATP and is expected to reach a saturation limit. In \cite{Kumar08} this shortcoming of the model has been explained by the unbounded spring potential, which is able to absorb an infinite amount of energy. Since our exact result for the velocity is valid for arbitrary potentials $U(x)$, an investigation of more realistic spring potentials, as for example a finitely extensible nonlinear elastic spring (FENE), is possible and presented in the following.

The FENE spring force is usually written as
\begin{eqnarray}
F(x)=-\frac{\kappa x}{1-(x/x_{m})^2},
\end{eqnarray}
where $\kappa$ denotes the spring constant and $x_m$ is the maximal extension of the spring. For small extensions $x$, the FENE spring behaves like a linear Hookean spring. The restoring force is rapidly enhanced for increased extension and becomes infinite in the limit $x\rightarrow x_m$ thus modelling the more realistic scenario of a finite extension. The corresponding potential is
\begin{eqnarray}
\label{I_FENE}
U(x)=-\frac{\kappa x_m^2}{2}\ln\left\{1-\left(\frac{x}{x_m}\right)^2\right\}.
\end{eqnarray}
For the FENE potential the distribution of relative coordinate and the average inchworm velocity can be determined as before, via Eqs.~(\ref{solution}) and (\ref{distribution}), where simply $U(x)$ has to be specified by Eq.~(\ref{I_FENE}). In order to obtain explicit expressions it is necessary to perform the integrals numerically. We use the same form of the stretch dependent damping, Eq.~(\ref{I_fric_vj}), and focus on two asymmetric scenarios, namely (i) asymmetric friction and equal noise strengths and (ii) equal friction on both beads and asymmetric noise.

\begin{figure}[htbp]
\begin{center}
\begin{tabular}{ll}
(a)\includegraphics[width=6.4cm]{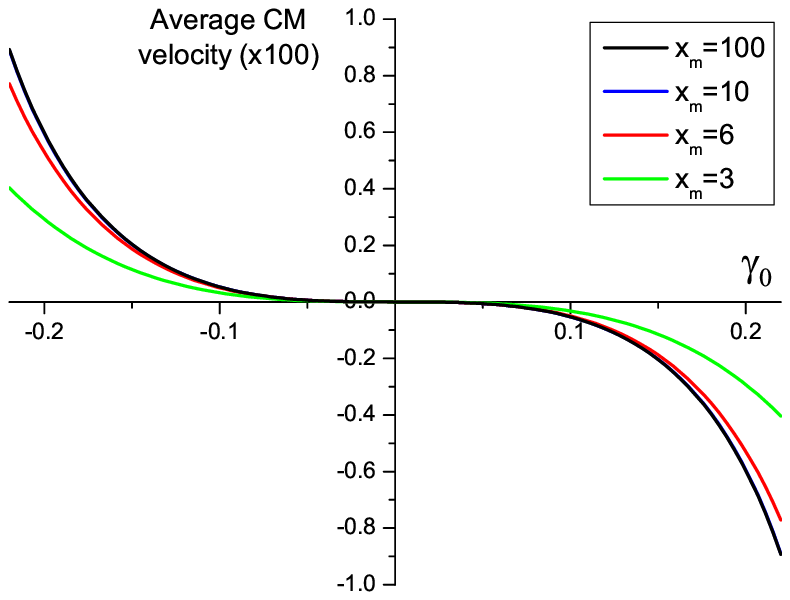} & (b)\includegraphics[width=6.4cm]{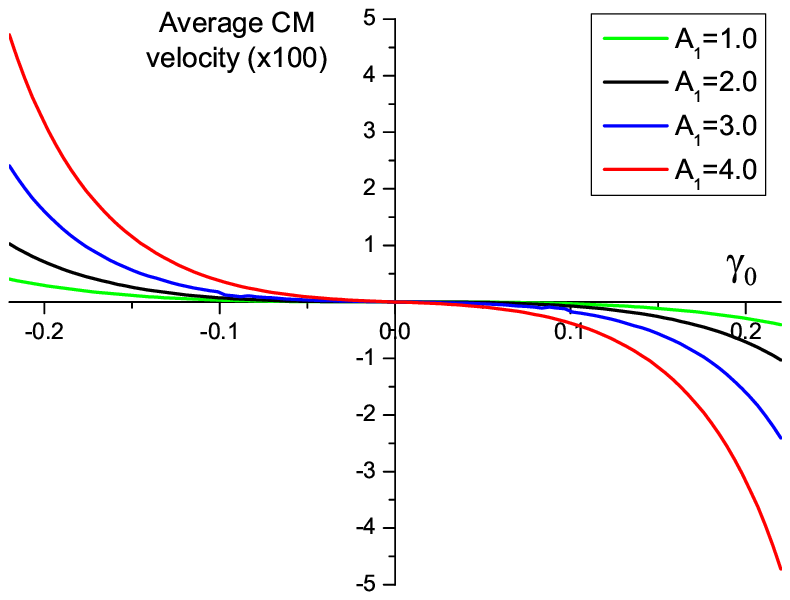}
\end{tabular}
\caption[Average inchworm velocity for the FENE spring in the asymmetric friction case.]{\label{I_Fig_FENEasymf}Average inchworm velocity for the FENE spring in the asymmetric friction case as a function of $\gamma_0$. (a) Noise strengths $A_1=A_2=1.0$ and four different values of $x_m$. (b) Fixed maximal extension at $x_m=3.0$ and increasing symmetric noise strengths $A_1=A_2$. Parameter values: $\kappa=0.05$, $w=4.0$.}
\end{center}
\end{figure}

\begin{figure}[htbp]
\begin{center}
\begin{tabular}{ll}
(a)\includegraphics[width=6.4cm]{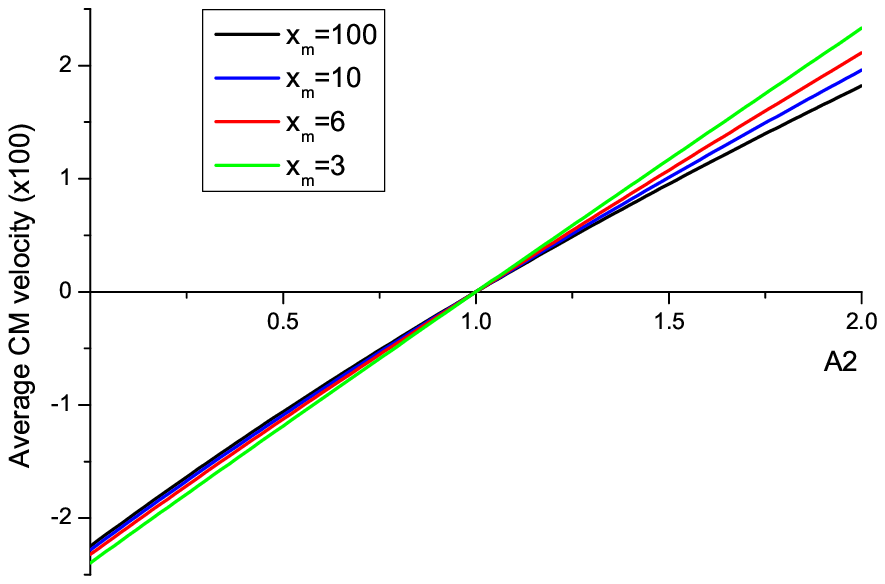}  & (b)\includegraphics[width=6.4cm]{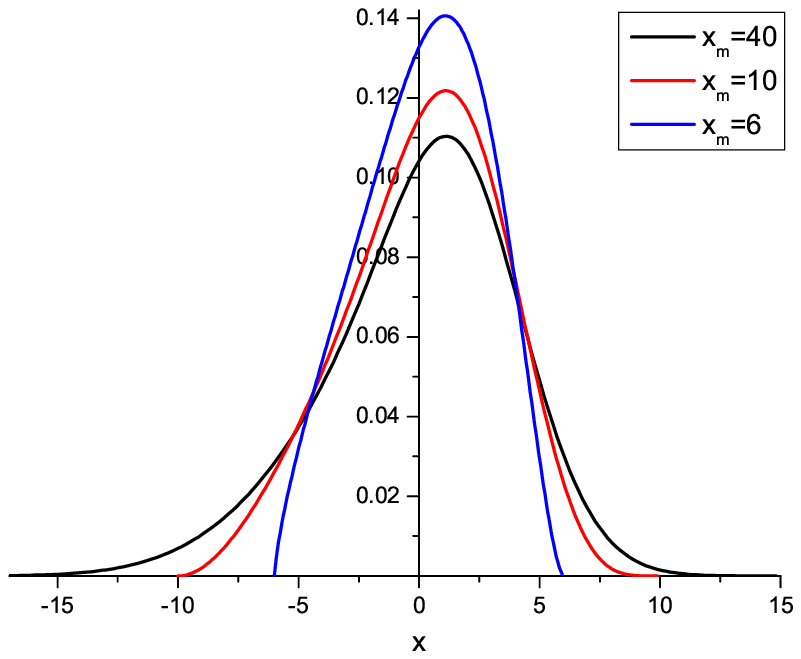}
\end{tabular}
\caption[The inchworm with FENE spring potential in the symmetric friction case.]{\label{I_Fig_FENEasymn}The inchworm with FENE spring potential in the symmetric friction case. (a) Average inchworm velocity as a function of $A_2$, where $A_1=1.0$ and $\gamma_0=0.1$. (b) Distribution of the relative coordinate for $A_1=1.0$ and $A_2=2.0$. Parameter values: $\kappa=0.05$, $w=4.0$, $\gamma_0=0.1$.}
\end{center}
\end{figure}

Fig.~\ref{I_Fig_FENEasymf}(a) shows results for the average inchworm velocity in case (i). We can see that up to $x_m=10$ the behaviour of the quadratic spring potential is reproduced. Constraining the extension of the spring further, noticeably reduces the inchworm velocity. On the other hand, if the noise strength on each bead is equally increased (see Fig.~\ref{I_Fig_FENEasymf}(b)), even for small maximal extension the velocity is considerably boosted. In the symmetric friction case (ii) a similar observation is made in Fig.~\ref{I_Fig_FENEasymn}(a), where the velocity is plotted as a function of $A_2$. Here the inchworm moves faster for smaller $x_m$. In all cases considered the qualitative features of the average inchworm velocity are very similar to the quadratic spring potential. A saturation of the velocity is not observed. Fig.~\ref{I_Fig_FENEasymn}(b) shows a plot of the distribution $p(x)$ in the symmetric friction case. For $x_m<10$ the FENE spring potential leads to an abrupt decay of the distribution in the vicinity of the maximal extension.

\section{Summary}
\label{I_Sec_over}

We have investigated a Brownian inchworm model of a self-propelled particle consisting of an elastic dimer driven by thermal and nonequilibrium noise. As our main results we derived exact expressions for the distribution of the relative coordinate and the average inchworm velocity for nonequilibrium Gaussian noise. The crucial property of the model allowing for an exact solution is the decoupling of the equations of motion of the center of mass and relative coordinate in the overdamped regime. The distribution of the relative coordinate is obtained in a straightforward way by solving the corresponding Fokker-Planck equation under the condition of stationarity and zero probability flux. For the average velocity it is necessary to determine noise averages over functionals of the noise, which can be calculated with the help of Novikov's theorem for Gaussian processes.

The exact solution shows excellent agreement with results from a direct simulation of the equations of motion retaining inertia and provides the exact foundation for some of the perturbative results in \cite{Kumar08}. For a FENE spring potential the behavior of the inchworm does not change significantly compared with a harmonic potential and in particular the velocity does not saturate. An inchworm-type model on a discrete lattice has been discussed in \cite{Kolomeisky05}, in which the mechanisms of walking and inchworming are
compared and criteria offered to distinguish which of these is operating in an
experimental system. The internal states in this model are discrete and finite
in number. It would be interesting to know whether the model, unlike ours,
shows saturation in the velocity as a function of an appropriately defined
parameter corresponding to ATP concentration.

Our Brownian inchworm exhibits a variety of different movement mechanisms depending on the stretch dependent friction, the noise strengths on each bead and possibly the spring potential and statistics of the noise. So far only the case of Gaussian nonequilibrium noise has been thoroughly examined. Future work should include noise with non-Gaussian statistics, for example Poissonian shot noise, which might be a more appropriate way to model chemical energy input. In a regime where the frequency of stochastic kicks is low compared with the dimer relaxation rate, the inchworm can fully relax in between noise inputs and might reveal a qualitative different behavior. However, the treatment would then require different analytical methods and is left for future work.

\acknowledgements

AB was funded under EPSRC Grant No. GR/T24593/01 while at the University of
Leeds and gratefully acknowledges the hospitality of the Indian Institute of
Science, Bangalore, where parts of this work were performed. SR acknowledges
support from CEFIPRA grant 3504-2 and the DST, India.

\begin{appendix}

\section{Adiabatic elimination}
\label{appendix}

In this appendix we perform the adiabatic elimination of the velocity for position-dependent friction. This derivation closely follows \cite{Ryter}.

Consider the following Langevin equation for a Brownian particle
\begin{eqnarray}
\dot{x} &=& v,
\\
m \dot{v} &=& -U'(x) - \gamma(x) v + g(x) \zeta(t),
\end{eqnarray}
with $\langle \zeta(t) \rangle  = 0; \,\, \langle \zeta(0) \zeta(t) \rangle = \delta(t)$ 
and the prime denotes differentiation w.r.t. $x$. 
The Fokker-Planck equation associated with the above Langevin equation is
\begin{equation}
 \frac{\partial P(x,v,t)}{\partial t} = -v \frac{\partial P(x,v,t)}{\partial x} + 
		\frac{1}{m} \frac{\partial}{\partial v} \Big[ U'(x) + \gamma(x) v + \frac{\partial}{\partial v} \frac{[g(x)]^2}{2m}  \Big] P(x,v,t).
\end{equation}

Defining
\begin{equation}
 Q_k (x,t) = \int dv \, v^k \, P(x,v,t)
\end{equation}
we need to find the evolution equation for $Q_o(x,t)$. It is easily shown that
\begin{eqnarray}
\frac{\partial Q_o}{\partial t} &=& -\frac{\partial Q_1 }{\partial x},
\\
\frac{\partial Q_1 }{\partial t} &=& -\frac{\partial Q_2 }{\partial x} - \frac{\gamma(x)}{m} Q_1 - \frac{U'(x)}{m} Q_o,
\\
\frac{\partial Q_2 }{\partial t} &=& -\frac{\partial Q_3 }{\partial x} - \frac{2 \gamma(x)}{m} Q_2 - \frac{2 U'(x)}{m} Q_1  
			+ \Big[ \frac{g(x)}{m} \Big]^2 Q_o.
\end{eqnarray}
The above equations are exact. We now take the overdamped limit (i.e., $m / \gamma(x) \ll \partial / \partial t$)
\begin{eqnarray}
Q_1 & \approx & -\frac{m}{\gamma(x)} \frac{\partial Q_2 }{\partial x} - \frac{U'(x)}{\gamma(x)}  Q_o,
\\
Q_2 & \approx & -\frac{m}{2\gamma(x)} \frac{\partial Q_3 }{\partial x} - \frac{U'(x)}{\gamma(x)} Q_1 
+ \frac{[g(x)]^2}{2m\gamma(x)}  Q_o.
\end{eqnarray}
Using the second equation above in the first, we get
\begin{eqnarray}
 Q_1 & = & \frac{m}{\gamma(x)} \frac{\partial}{\partial x} \Bigg( \frac{m}{2\gamma(x)} \frac{\partial Q_3 }{\partial x} 
	+ \frac{U'(x)}{\gamma(x)} Q_1 - \frac{[g(x)]^2}{2m\gamma(x)}  Q_o \Bigg)
	- \frac{U'(x)}{\gamma(x)} Q_o
\nonumber \\
& \approx & - \frac{1}{2\gamma(x)} \frac{\partial}{\partial x} \frac{[g(x)]^2}{\gamma(x)}  Q_o 	- \frac{U'(x)}{\gamma(x)} Q_o
\end{eqnarray}
to lowest order in $m / \gamma(x)$. Thus
\begin{eqnarray}
\frac{\partial Q_o (x,t)}{\partial t} &=& \frac{\partial}{\partial x} 
		\Bigg( \frac{ U'(x)}{\gamma(x)} + \frac{1}{2\gamma(x)} \frac{\partial}{\partial x} \frac{[g(x)]^2}{\gamma(x)} \Bigg) Q_o(x,t)
\nonumber \\
		&=& \frac{\partial}{\partial x} 
		\Bigg( \frac{U'(x)}{\gamma(x)} + \frac{\gamma'(x) [g(x)]^2 }{2[\gamma(x)]^3} 
			+ \frac{1}{2} \frac{\partial}{\partial x} \bigg[ \frac{g(x)}{\gamma(x)} \bigg]^2 \Bigg) Q_o(x,t).
\end{eqnarray}
This implies the following Langevin equations
\begin{eqnarray}
\dot{x} &=& -\frac{U'(x)}{\gamma(x)} - \frac{\gamma'(x) [g(x)]^2}{2 [\gamma(x)]^3} + \frac{g(x)}{\gamma(x)} \eta(t),
\qquad {\rm It \hat{o} }
\\
 \dot{x} &=& -\frac{U'(x)}{\gamma(x)} - \frac{g(x) g'(x)}{2 [\gamma(x)]^2} \quad + \frac{g(x)}{\gamma(x)} \eta(t),
\qquad {\rm Stratonovich}
\end{eqnarray}
where $\eta$ is a zero mean Gaussian white noise.

\end{appendix}


\begin{thebibliography}{}
\expandafter\ifx\csname natexlab\endcsname\relax\def\natexlab#1{#1}\fi
\expandafter\ifx\csname bibnamefont\endcsname\relax
  \def\bibnamefont#1{#1}\fi
\expandafter\ifx\csname bibfnamefont\endcsname\relax
  \def\bibfnamefont#1{#1}\fi
\expandafter\ifx\csname citenamefont\endcsname\relax
  \def\citenamefont#1{#1}\fi
\expandafter\ifx\csname url\endcsname\relax
  \def\url#1{\texttt{#1}}\fi
\expandafter\ifx\csname urlprefix\endcsname\relax\def\urlprefix{URL }\fi
\providecommand{\bibinfo}[2]{#2}
\providecommand{\eprint}[2][]{\url{#2}}

\bibitem[{\citenamefont{Howard}(2001)}]{Howard}
\bibinfo{author}{\bibfnamefont{J.}~\bibnamefont{Howard}},
  \emph{\bibinfo{title}{{Mechanics of Motor Proteins and the Cytoskeleton}}}
  (\bibinfo{publisher}{Sinauer Associates, Sunderland}, \bibinfo{year}{2001}).

\bibitem[{\citenamefont{Yu et~al.}(2006)\citenamefont{Yu, Ha, and
  Schulten}}]{Yu06}
\bibinfo{author}{\bibfnamefont{J.}~\bibnamefont{Yu}},
  \bibinfo{author}{\bibfnamefont{T.}~\bibnamefont{Ha}}, \bibnamefont{and}
  \bibinfo{author}{\bibfnamefont{K.}~\bibnamefont{Schulten}},
  \bibinfo{journal}{Biophysical Journal} \textbf{\bibinfo{volume}{91}},
  \bibinfo{pages}{2097} (\bibinfo{year}{2006}).

\bibitem[{\citenamefont{Altman et~al.}(2004)\citenamefont{Altman, Sweeney, and
  Spudich}}]{Altman04}
\bibinfo{author}{\bibfnamefont{D.}~\bibnamefont{Altman}},
  \bibinfo{author}{\bibfnamefont{H.~L.} \bibnamefont{Sweeney}},
  \bibnamefont{and} \bibinfo{author}{\bibfnamefont{J.~A.}
  \bibnamefont{Spudich}}, \bibinfo{journal}{Cell}
  \textbf{\bibinfo{volume}{116}}, \bibinfo{pages}{737} (\bibinfo{year}{2004}).

\bibitem[{\citenamefont{Yamada et~al.}(2003)\citenamefont{Yamada, Hondou, and
  Sano}}]{Yamada03}
\bibinfo{author}{\bibfnamefont{D.}~\bibnamefont{Yamada}},
  \bibinfo{author}{\bibfnamefont{T.}~\bibnamefont{Hondou}}, \bibnamefont{and}
  \bibinfo{author}{\bibfnamefont{M.}~\bibnamefont{Sano}},
  \bibinfo{journal}{Physical Review E} \textbf{\bibinfo{volume}{67}},
  \bibinfo{pages}{40301} (\bibinfo{year}{2003}).

\bibitem[{\citenamefont{Dorbolo et~al.}(2005)\citenamefont{Dorbolo, Volfson,
  Tsimring, and Kudrolli}}]{Dorbolo05}
\bibinfo{author}{\bibfnamefont{S.}~\bibnamefont{Dorbolo}},
  \bibinfo{author}{\bibfnamefont{D.}~\bibnamefont{Volfson}},
  \bibinfo{author}{\bibfnamefont{L.}~\bibnamefont{Tsimring}}, \bibnamefont{and}
  \bibinfo{author}{\bibfnamefont{A.}~\bibnamefont{Kudrolli}},
  \bibinfo{journal}{Physical Review Letters} \textbf{\bibinfo{volume}{95}},
  \bibinfo{pages}{044101} (\bibinfo{year}{2005}).

\bibitem{julicheretal} F. J\"ulicher, A. Ajdari, and J. Prost, Reviews of Modern Physics \textbf{69}, 1269 (1997).

\bibitem[{\citenamefont{Reimann}(2002)}]{Reimann02}
\bibinfo{author}{\bibfnamefont{P.}~\bibnamefont{Reimann}},
  \bibinfo{journal}{Physics Reports} \textbf{\bibinfo{volume}{361}},
  \bibinfo{pages}{57} (\bibinfo{year}{2002}).

\bibitem{Kolomeisky07}A. B. Kolomeisky and M. E. Fisher, Annu. Rev. Phys. Chem. \textbf{58}, 675 (2007). 

\bibitem[{\citenamefont{Mogilner et~al.}(1998)\citenamefont{Mogilner, Mangel,
  and Baskin}}]{Mogilner98}
\bibinfo{author}{\bibfnamefont{A.}~\bibnamefont{Mogilner}},
  \bibinfo{author}{\bibfnamefont{M.}~\bibnamefont{Mangel}}, \bibnamefont{and}
  \bibinfo{author}{\bibfnamefont{R.~J.} \bibnamefont{Baskin}},
  \bibinfo{journal}{Physics Letters A} \textbf{\bibinfo{volume}{237}},
  \bibinfo{pages}{297} (\bibinfo{year}{1998}).

\bibitem[{\citenamefont{Kumar et~al.}(2008)\citenamefont{Kumar, Ramaswamy, and
  Rao}}]{Kumar08}
\bibinfo{author}{\bibfnamefont{K.~V.} \bibnamefont{Kumar}},
  \bibinfo{author}{\bibfnamefont{S.}~\bibnamefont{Ramaswamy}},
  \bibnamefont{and} \bibinfo{author}{\bibfnamefont{M.}~\bibnamefont{Rao}},
  \bibinfo{journal}{Physical Review E} \textbf{\bibinfo{volume}{77}},
  \bibinfo{pages}{020102} (\bibinfo{year}{2008}).

\bibitem[{\citenamefont{{Van Kampen}}(1992)}]{VanKampen}
\bibinfo{author}{\bibfnamefont{N.~G.} \bibnamefont{{Van Kampen}}},
  \emph{\bibinfo{title}{{Stochastic Processes in Physics and Chemistry}}}
  (\bibinfo{publisher}{North-Holland, Amsterdam}, \bibinfo{year}{1992}).
  
\bibitem{VanKampen2} N. G. Van Kampen, IBM J. Res. Develop, \textbf{32},
107 (1988); Phys Rep, \textbf{124}, 69 (1985).

\bibitem{Gardiner} C. W. Gardiner, Handbook of Stochastic Methods (2nd
edition), Springer (1985), esp. after eq. (6.4.11).

\bibitem{Ryter} D. Ryter,  Z. Physik B - Condensed Matter \textbf{41}, 39 (1981).

\bibitem[{\citenamefont{Sancho et~al.}(1982)\citenamefont{Sancho, Miguel, and
  D{\"u}rr}}]{Sancho82}
\bibinfo{author}{\bibfnamefont{J.~M.} \bibnamefont{Sancho}},
  \bibinfo{author}{\bibfnamefont{M.~S.} \bibnamefont{Miguel}},
  \bibnamefont{and} \bibinfo{author}{\bibfnamefont{D.}~\bibnamefont{D{\"u}rr}},
  \bibinfo{journal}{Journal of Statistical Physics}
  \textbf{\bibinfo{volume}{28}}, \bibinfo{pages}{291} (\bibinfo{year}{1982}).
  
\bibitem{Lau07} A. W. C. Lau and T. C. Lubensky, Phys. Rev. E \textbf{76}, 011123 (2007).

\bibitem[{\citenamefont{Novikov}(1965)}]{Novikov65}
\bibinfo{author}{\bibfnamefont{E.~A.} \bibnamefont{Novikov}},
  \bibinfo{journal}{Soviet Physics --- JETP} \textbf{\bibinfo{volume}{20}},
  \bibinfo{pages}{1290} (\bibinfo{year}{1965}).

\bibitem[{\citenamefont{Fox}(1986)}]{Fox86}
\bibinfo{author}{\bibfnamefont{R.~F.} \bibnamefont{Fox}},
  \bibinfo{journal}{Physical Review A} \textbf{\bibinfo{volume}{33}},
  \bibinfo{pages}{467} (\bibinfo{year}{1986}).

\bibitem[{\citenamefont{Risken}(1996)}]{Risken}
\bibinfo{author}{\bibfnamefont{H.}~\bibnamefont{Risken}},
  \emph{\bibinfo{title}{{The Fokker-Planck Equation: Methods of Solution and
  Applications}}} (\bibinfo{publisher}{Springer, Berlin},
  \bibinfo{year}{1996}).

\bibitem{kloeden}
P.~E. Kloden, E. Platen, and H. Schurz, {\em Numerical solution of SDE through
  Computer experiments} (Springer-Verlag, Berlin-Heidelberg, 1994).

\bibitem{ayushmansen}
W. F. Paxton, A. Sen, and T. E. Mallouk, Chem. Eur. J. 11, 6462 (2005)

\bibitem{ramin}
J. R. Howse, R. A. L. Jones, A. J. Ryan, T. Gough, R. Vafabakhsh, and R. Golestanian, 
Phys. Rev. Lett. 99, 048102 (2007).

\bibitem{kapral}
G. R\"{u}ckner and R. Kapral, Phys. Rev. Lett {\bf 98},  150603  (2007).

\bibitem{Kolomeisky05}A. B. Kolomeisky and H. Phillips III, J. Phys.: Condens. Matter \textbf{17}, S3887 (2005).

\end{thebibliography}
\end{document}